\documentclass{optica-article}
\journal{opticajournal}
\articletype{Research Article}
\usepackage{lineno}
\usepackage{algorithm}
\usepackage{algpseudocode}
\usepackage{float}

\begin{document}

\title{Grover-Sagnac interferometer}

\author{Christopher R. Schwarze,\authormark{1, *} Anthony D. Manni,\authormark{1} David S. Simon,\authormark{1,2} Abdoulaye Ndao,\authormark{3} and Alexander V. Sergienko\authormark{1,4}}
\address{\authormark{1}Department of Electrical and Computer Engineering \& Photonics Center, Boston University, 8 Saint Mary’s St., Boston, Massachusetts 02215, USA\\
\authormark{2}Department of Physics and Astronomy, Stonehill College, 320 Washington Street, Easton, Massachusetts 02357, USA\\
\authormark{3}Department of Electrical and Computer Engineering, University of California, San Diego, La Jolla, California 92093-0401, USA\\
\authormark{4}Department of Physics, Boston University, 590 Commonwealth Avenue, Boston, Massachusetts 02215, USA}
\email{\authormark{*}crs2@bu.edu}

\begin{abstract}
We demonstrate a nontraditional design of the Sagnac interferometer by replacing the commonly used beam splitter with a linear-optical Grover multiport. This substitution creates a pole at the origin of the device parameter space with an associated resonance in the output intensity. The structure of this resonance is dictated only by the non-reciprocal portion of the phase acquired in the Sagnac loop. This property directly results from adopting the more symmetric and higher-dimensional central scattering coin, and allows for a different approach to registering and detecting the non-reciprocal Sagnac phase. This parameter may be extracted from the width of a peak or dip in the interferogram instead of tracing small changes in power as in traditional Sagnac interferometry. We discuss how losses affect the system and potential metrological applications.
\end{abstract}

\section{Introduction}

Interferometric systems dating back to early days of electromagnetism remain prevalent today in part due to their simple, low-dimensional parameter spaces and minimal number of required resources. Together these characteristics underlie robust optical systems typically operating by two mode interference or a basic resonant cavity, as exemplified by the Michelson, Mach-Zehnder, and Fabry-Pérot configurations. These devices are all comprised of traditional beam-splitting scatterers, which divide and recombine two forward moving field modes according to a beam-splitter transformation, 
\begin{equation}\label{eq:staticu}
U = 
\begin{pmatrix}    
    r_1 & t_2\\
    t_1 & r_2
\end{pmatrix}.
\end{equation}
Since $U$ is a $2\times 2$ matrix, the beam-splitters which are mathematically embodied by it are classified as two-dimensional linear optical scatterers. The physical implementation of such a scattering device is otherwise arbitrary. To name but a few examples, this matrix model can be closely realized by a cube beam-splitter formed by two connected prisms, a half-silvered mirror, or a stack of dielectric layers deposited on a transparent substrate. The scattering amplitudes $r_j$ and $t_j$ and thus $U$ itself will generally vary with attributes of the underlying field modes such as polarization, frequency and/or waveguide propagation constant. The present analysis will be confined to monochromatic, linearly polarized radiation so that dispersive effects due to this variation can be ignored. When this is done, the matrix of Eq. (\ref{eq:staticu}) may be regarded as a static, or passive device; any tunable parameter such as temperature or strain that could modify it are assumed to be fixed. 

An interferometer is a tunable device comprised of one or more passive scatterers. The aggregated collection of scattering units may be embodied by a single scattering matrix that depends on one or more physical parameters, which are often given by a change in the optical phase $\phi = 2\pi n \ell/\lambda$. Tuning $\phi$ through variations in path-length $\ell$ or wavelength $\lambda$ are both standard practices, as are a myriad of additional approaches enabled by perturbative changes to the refractive index $n$, such as the thermo-optic, acousto-optic and electro-optic effects.

Higher-dimensional optical systems redistribute energy among more than two field modes. Because of this, the simpler low-dimensional systems mentioned above are comparatively limited in some regards. Higher-dimensional systems are naturally equipped for higher information throughput, and beyond that, can exhibit a broader range of interference behavior. This has led to various proposals and realizations of higher-dimensional interferometric systems with pronounced advantages over their low-dimensional counterparts. 
\begin{figure}[ht]
    \centering
    \includegraphics[width=.5\linewidth]{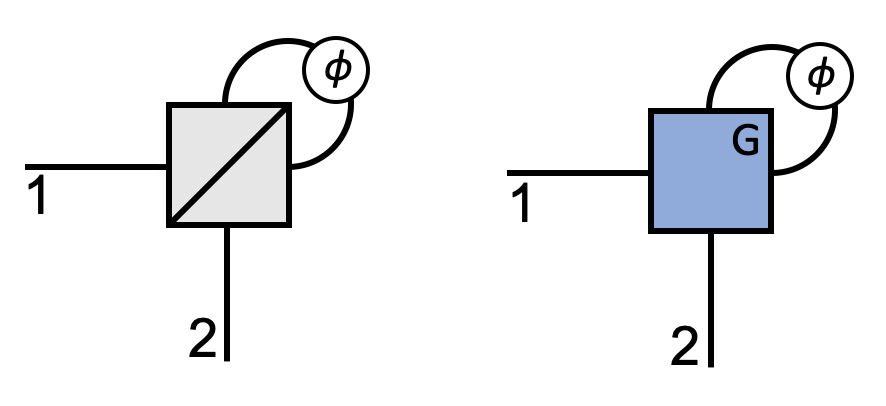}
    \caption{(left) Traditional Sagnac interferometer. (right) Grover-Sagnac interferometer, formed with a Grover four-port (Eq. \ref{eq:grover}) instead of a beam splitter.}
    \label{fig:sagnacs}
\end{figure}
An early class of such systems is based on the notion of light recycling, which was originally investigated as a means to improve gravitational wave interferometers \cite{PhysRevD.38.2317, Fritschel:92, PhysRevLett.66.1391} and ultimately became a critical aspect of the present design at the Laser Interferometer Gravitational-Wave Observatory (LIGO) \cite{LIGO}. In light recycling systems, light from an output port is fed back into the device, creating internally coupled resonators that can increase the output signal-to-noise ratio. Light recycling was originally viewed as a counterintuitive phenomenon because a source of loss (a partial reflector) would be placed in front of the detector, yet the signal-to-noise ratio would increase. The effect is a consequence of increasing the modal dimensionality of the interference. Generally, coherently coupled resonators will behave effectively like a single, one-dimensional resonator which possesses now-tunable properties that are normally intrinsic to the materials used to form a regular uncoupled optical cavity. A recurring example is the cavity finesse; although derived from the mirror reflectivities in a traditional linear cavity, in coupled-cavity systems, the finesse can be phase tunable \cite{Gray:98, PhysRevA.107.052615}.

Light recycling systems constructed from two-dimensional beam-splitters can also be viewed as lying within a subset of a broader class of multimode interferometers. Using higher-dimensional multiport scatterers in place of beam splitters can offer additional advantages. Resonator-free multiport interferometers serve as a comparatively stable means of increasing the phase response or to access information about multiple phase shifts simultaneously \cite{Weihs:96, PhysRevA.106.033706}. Meanwhile, in coupled resonator systems, higher-dimensional scattering coins can realize especially useful parametrizations of the system phase response. 

An important example is the four-port Grover coin, which enacts the following passive transformation:
\begin{equation}\label{eq:grover}
G = \frac12
\begin{pmatrix}
     -1 & 1 & 1 & 1\\
     1 & -1 & 1 & 1\\
     1 & 1 & -1 & 1\\
     1 & 1 & 1 & -1
\end{pmatrix}.
\end{equation}
This device can replace the traditional four-port beam-splitter, augmenting its dimensionality in the process. Several ways of implementing this passive scatterer are known, including an experimentally demonstrated, low-resource Y-coupler decomposition \cite{PhysRevA.110.023527} that forms the coin without relying on interference, making it more stable than traditional beam-splitter mesh decompositions of multiports \cite{Clements:16}. 

This Grover multiport is also more symmetric than the beam-splitter, since input to any port produces the same output field pattern. Another key characteristic of this device is that, unlike the beam-splitter, it does not possess a directional bias. A photon entering one of its ports has a $1/4$ probability of back-reflecting, as well as a $1/4$ probability for emerging at each of the three other ports. Collectively, linear scattering devices with an intended, coherent back-reflection like this are known as directionally-unbiased. A more common example is a dielectric slab operating at normal incidence, which possesses a scattering matrix in the form of Eq. (\ref{eq:staticu}) with $r_j$ and $t_j$ given by the Fresnel coefficients. A diffraction grating operating in the Littrow configuration also functions as an unbiased multiport.

The Y-coupler derived form of a Grover four port was recently used to demonstrate a tunable slope enhancement in a Grover-Michelson interferometer \cite{Schwarze:24}. Replacing the traditional beam-splitter with the Grover coin converts the Michelson interferometer arms to coupled resonators, causing a parametric redundancy to be removed: in a traditional Michelson interferometer, one arm phase $\phi_1$ defines an interferogram curve $P$ while the other arm phase $\phi_2$ translates this curve, because $P$ is only a function of the quantity $(\phi_1 - \phi_2)$. In contrast, the Grover-Michelson nonlinearly mixes the phase parameters $\phi_1, \phi_2$ in a manner that skews the traditional sinusoidal Michelson interferogram when $\phi_2$ is varied. This parametrization alters the maximum slope of the interferogram with respect to $\phi_1$ while leaving the visibility unchanged, allowing the system sensitivity and stability to be exchanged much like a zoom lens allows magnification to be traded with field of view.

In this work, we now consider the Grover-Sagnac interferometer \cite{PhysRevA.107.052615}, which benefits in its own manner from the increased dimensionality and symmetry of the four-port Grover coin. The increase in dimensionality again lifts a parametric redundancy while the symmetry underlies the property that the device responds resonantly to only the non-reciprocal portions of the loop phase. In other words, without a non-reciprocal phase, the device is a single-pass system, but with such a term, the device is elevated to a coupled resonator system. Together, the freed parameter and selectively non-reciprocal resonance lead to a direct and stable readout for the non-reciprocal phase that differs from prior approaches to resonant Sagnac systems \cite{Shaddock:98, Liu:23}.

The Grover-Sagnac system and readout will be discussed in further detail in Section III, after a brief review of the traditional Sagnac interferometer and associated Sagnac effect in Section II. In Section IV the effects of optical loss on each system are considered. This analysis automatically extends to optical gain in the Sagnac loop, since gain may be represented by a negative loss parameter. In Section V, some metrological applications of the presented system are discussed before drawing final conclusions in Section VI.

\section{Sagnac interferometry}

A traditional Sagnac interferometer is formed by replacing the mirrors of a traditional Michelson interferometer with a loop, resulting in the arrangement shown in Fig. \ref{fig:sagnacs} (left). Light enters the system from port 1 or 2, impinging a beam-splitter which then evenly divides the incident light into clockwise and counterclockwise components traversing the loop. The common-path configuration ensures that the loop excitations acquire the same phase $\phi$. 

In some special circumstances, counter-propagating field modes of the same wavelength and polarization can acquire different phases, marking a local violation of reciprocity. A common example of such circumstances associated with the Sagnac interferometer configuration is the Sagnac effect \cite{RevModPhys.39.475}: when the entire device is rotated, the round-trip distance of the loop is reduced in the direction of rotation relative to the opposing direction while the speed of propagation of the light remains unchanged. If the angular velocity vector of the rotating apparatus is $\mathbf{\Omega}$, and the loop area vector is $\mathbf{A}$, then the rotation results in a non-reciprocal phase delay between the two counter-propagating excitations of 
\begin{equation}\label{eq:sagnac-phase}
    \Delta \approx \bigg (\frac{8\pi}{\lambda c} \bigg ) \mathbf{A}\cdot \mathbf{\Omega}.
\end{equation}
This approximation holds for $v \ll c$, where $v$ is the magnitude of the velocity of the beam-splitter relative to a stationary observer, and $c$ is the vacuum speed of light. So, to maximize $\Delta$ at a fixed wavelength, the area and angular velocity vectors should be collinear and with maximum magnitudes. 

The Sagnac configuration in Fig. \ref{fig:sagnacs} (left) is perhaps the simplest method of determining a non-reciprocal phase delay, $\Delta$, whether it is produced by the Sagnac effect or some other method, such as a time-varying stress \cite{Liu:23}, Faraday rotator, etc. Measuring $\Delta$ with this device allows a physical quantity such as angular velocity to be inferred directly. This metrological application has been substantially developed for modern navigation systems. With a sufficiently large loop area $|\mathbf{A}|$, very small angular velocities are able to induce a sizable $\Delta$. This approach enabled the rotation of the earth to be detected by Michelson and Gale in 1925 \cite{MICHELSON1925}. Contemporary systems employ a tightly coiled fiber loop to obtain a large effective area and reduced signal loss with a compact device. Many modern variations of the original Sagnac interferometer have been developed for applications in thin-film measurement \cite{Usman24_2, Usman_2022}, temperature and strain sensing \cite{10597582}, and general phase metrology \cite{Usman24_1}.

To find the output of a traditional Sagnac interferometer, we will assume the beam-splitter in Fig. 1 (left) scatters light according to the transformation 
\begin{equation}\label{eq:bs}
\frac{1}{\sqrt{2}}
\begin{pmatrix}
    i & 1\\
    1 & i
\end{pmatrix}.
\end{equation} 
Photons propagating away from ports 1 or 2 will be respectively identified with the photon creation operators $a_1^\dagger$ and $a_2^\dagger$, while that for (counter)clockwise traveling photons in the loop will be $a_{\text{(c)cw}}^\dagger$. This notation generally applies to nonclassical optical states, but in the present single-photon analysis, the operators acting on vacuum may be equivalently replaced by the amplitude of a classical coherent field, e.g. $a_1^\dagger|0\rangle \rightarrow E_1$, etc.

For an initial state of a photon impinging port 1, the beam-splitter maps the state to $(i a_{\text{cw}}^\dagger + a_{\text{ccw}}^\dagger)/\sqrt{2} |0\rangle$. After traversing the loop, this becomes $e^{i\phi}(i e^{i\delta}a_{\text{cw}}^\dagger + e^{-i\delta} a_{\text{ccw}}^\dagger)/\sqrt{2} |0\rangle$ where $\delta = \Delta/2$. Evenly distributing the non-reciprocal phase $\Delta$ over both modes like this is equivalent to applying it entirely to one mode or redistributing it arbitrarily over both; such a redistribution of $\delta$ is easily seen to be equivalent to a translation in the value of $\phi$. This translation can be discarded because a global value of $\phi$ can not be determined anyways. When the state revisits the beam-splitter, it is transformed to the output state 
\begin{subequations}
\begin{align}
    |\psi_f^{(1)}\rangle &= \frac{e^{i\phi}}{2} \bigg (i e^{i\delta}(a_1^\dagger + i a_2^\dagger) + e^{-i\delta} (i a_1^\dagger + a_2^\dagger ) \bigg ) |0\rangle\\
    &= i e^{i\phi} (\cos\delta a_1^\dagger - \sin\delta a_2^\dagger ) |0\rangle.
\end{align}
\end{subequations}
A similar analysis for input to port 2 reveals the corresponding output state to be 
\begin{equation}
    |\psi_f^{(2)}\rangle = i e^{i\phi}  (\sin\delta a_1^\dagger + \cos\delta a_2^\dagger) |0\rangle.
\end{equation}
These results may then be combined to form an active device scattering matrix $U_S(\phi, \delta = \Delta/2)$ for the Sagnac interferometer,
\begin{equation}\label{eq:sagnacS}
    U_{\text{S}}(\phi, \delta) = i e^{i\phi} 
    \begin{pmatrix}
        \cos\delta & \sin\delta\\
        -\sin\delta & \cos\delta
    \end{pmatrix}.
\end{equation}
To first order in $\delta$ the magnitude of the diagonal entries of this matrix are one, indicating that for small $\delta$ this device behaves like a mirror for all values of $\phi$. This is the underlying operating principle of a Sagnac loop mirror, which is a Sagnac interferometer in the basic configuration of Fig. \ref{fig:sagnacs} (left) operated passively, without any need to tune the loop phase. It has become a standard on-chip photonic component \cite{10.1063/5.0123236}. In these systems, the loop area is so small that even sizable rotations can be neglected. 

We see the device $U_\text{S}$ produces a sinusoidal interferogram with transmission probability $T(\delta) = \sin^2(\delta)$. This is the same interferogram shape produced by a Michelson or Mach-Zehnder interferometer except that only the non-reciprocal phase traces out the curve; the loop phase $\phi$ has no effect on the probabilities, making the interferogram and produced fringes comparatively more stable. Non-reciprocal phases can still have intrinsic fluctuations nonetheless: in the Michelson-Gale experiment \cite{MICHELSON1925}, time-dependent atmospheric fluctuations required the experiment to be carried out within an evacuated pipe.

In the traditional Sagnac interferometer, $\delta$ can be determined from a measurement of the ratio of transmitted power to incident power. While this is a simple, direct procedure, it is not especially sensitive for small $\delta$. Since the change in $T$ is zero to first order in $\delta$ about $\delta = 0$, very small changes in power must be detected, and these changes must be statistically significant relative to the intrinsic energy fluctuations in the source and detector subsystems.
\section{Grover-Sagnac interferometry}

A Grover-Sagnac interferometer can be formed by replacing the beam-splitter described by Eq. (\ref{eq:bs}) with the Grover coin defined in Eq. (\ref{eq:grover}), as is depicted in Fig. \ref{fig:sagnacs} (right). In the Grover-Michelson interferometer, this replacement generates two arm-resonators that are coupled to each other by the Grover coin. So it might be expected that the loop here would lead to a coupling between the counter-propagating ring resonator modes $a_{\text{cw}}^\dagger$ and $a_{\text{ccw}}^\dagger$. This does occur in general, but not when $\delta = 0$: after a photon impinges port 1, the Grover coin scatterers the photon into the state $(-a_1^\dagger + a_2^\dagger + a_{\text{cw}}^\dagger + a_{\text{ccw}}^\dagger)/2|0\rangle$. The symmetrically excited loop modes each acquire a round-trip phase $\phi$ and recombine at the Grover coin. Momentarily ignoring the initial excitations outside the loop, we see the loop-mode state portion $e^{i\phi} (a_{\text{cw}}^\dagger + a_{\text{ccw}}^\dagger)/2$ would then be mapped by the coin to $e^{i\phi} (a_1^\dagger + a_2^\dagger - a_{\text{cw}}^\dagger + a_{\text{ccw}}^\dagger)/4 + e^{i\phi} (a_1 ^\dagger + a_2^\dagger + a_{\text{cw}}^\dagger-a_{\text{ccw}}^\dagger)/4 = e^{i\phi} (a_1 ^\dagger + a_2^\dagger)/2$, extinguishing any recursive re-excitations inside the loop. Combining this with the initial portion, we see the final state is 
\begin{subequations}
\begin{align}
    |\psi_f^{(1)}\rangle &= \frac12 \bigg ( (e^{i\phi} - 1) a_1^\dagger + (e^{i\phi} + 1) a_2^\dagger \bigg ) |0\rangle \\ 
    &= e^{i\phi/2} (i \sin(\phi/2)a_1^\dagger + \cos(\phi/2) a_2^\dagger)|0\rangle\label{eq:gsi}.
\end{align}
\end{subequations}
The case of a photon input to port 2 can be viewed as a relabeling of the Grover coin ports $1 \leftrightarrow 2$. Since the Grover coin matrix of Eq. (\ref{eq:grover}) is invariant under any port relabeling, the relabeling process will not affect the resultant scattering dynamics. Therefore the output state is the same as Eq. (\ref{eq:gsi}), which after undoing the labeling, corresponds to a swap of the modal labels 1 and 2 in the above. This gives a final scattering matrix of
\begin{equation}\label{eq:stationary}
U_{\text{GS}}(\phi, \delta = 0) = e^{i\phi/2}
\begin{pmatrix}
    i\sin(\phi/2) & \cos(\phi/2)\\
    \cos(\phi/2) & i\sin(\phi/2)
\end{pmatrix}.
\end{equation}
The increased dimensionality of the Grover coin generates transient excitations outside the loop that interfere with the energy that later emerges from the loop, giving rise to the above $\phi$ dependence. A beam-splitter does not create any separate transient excitation; rather all of the energy enters and exits the loop together, so that the variable $\phi$ simply becomes a global phase shift to the entire state. As a result, $\phi$ does not affect the scattering probabilities in a conventional Sagnac interferometer.

In the Grover-Sagnac interferometer, with $\delta = 0$, tuning $\phi$ allows the device to be brought to any splitting ratio with sinusoidal dependence, acting as a proper one-parameter Michelson interferometer. Furthermore, the (anti)symmetric action of the coin on the loop modes ensures all recursive reflections back into the loop nullify if $\delta = 0$.

\begin{figure}[ht]
    \centering
    \includegraphics[width=.5\linewidth]{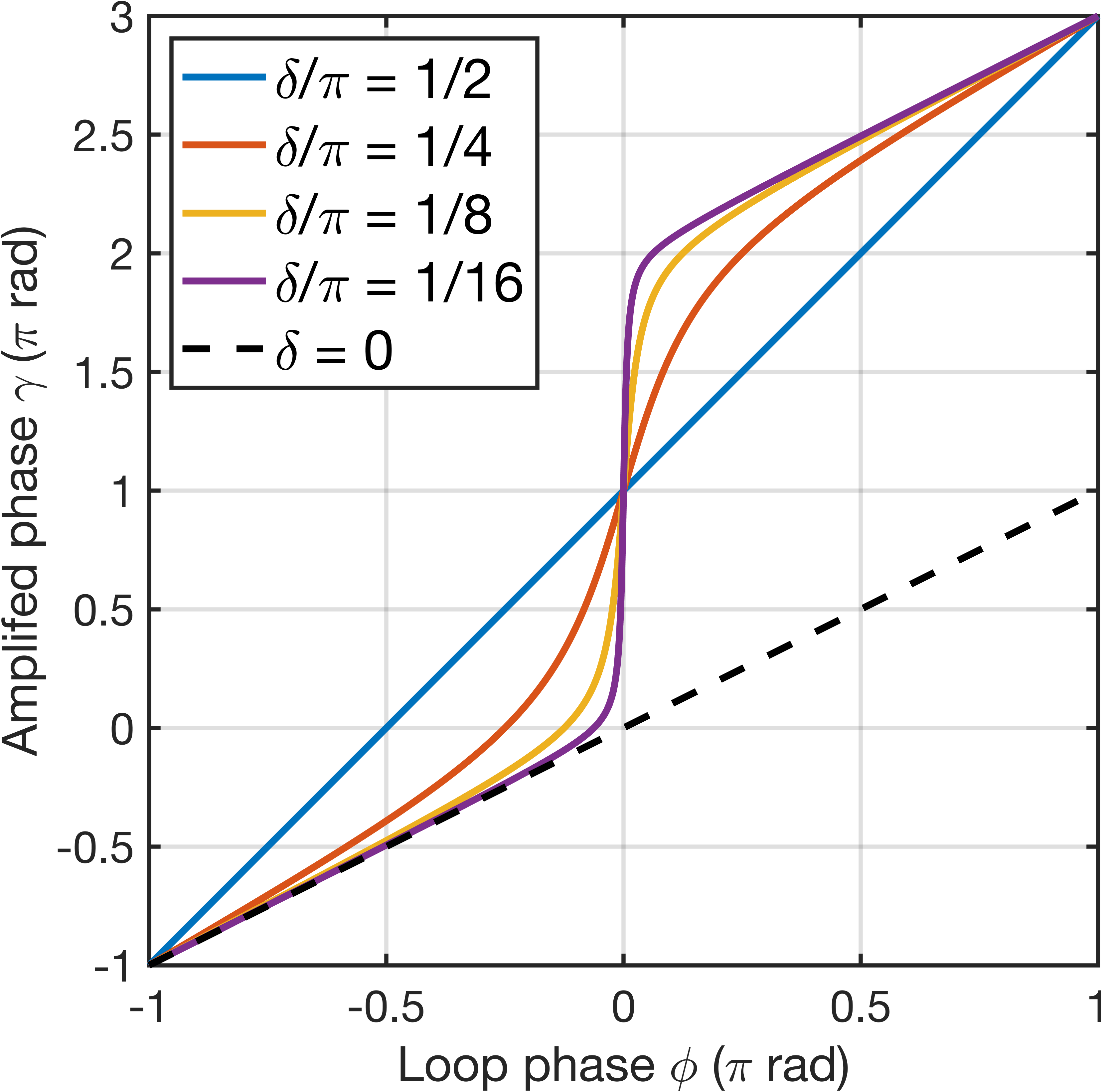}
    \caption{Generalized phase response $\gamma$ (Eq. (\ref{eq:gamma})) of the Grover-Sagnac interferometer loop as a function of the reciprocal portion of the loop phase $\phi$.}
    \label{fig:gammas}
\end{figure}

\begin{figure}[H] 
    \centering
    \includegraphics[width=.5\linewidth]{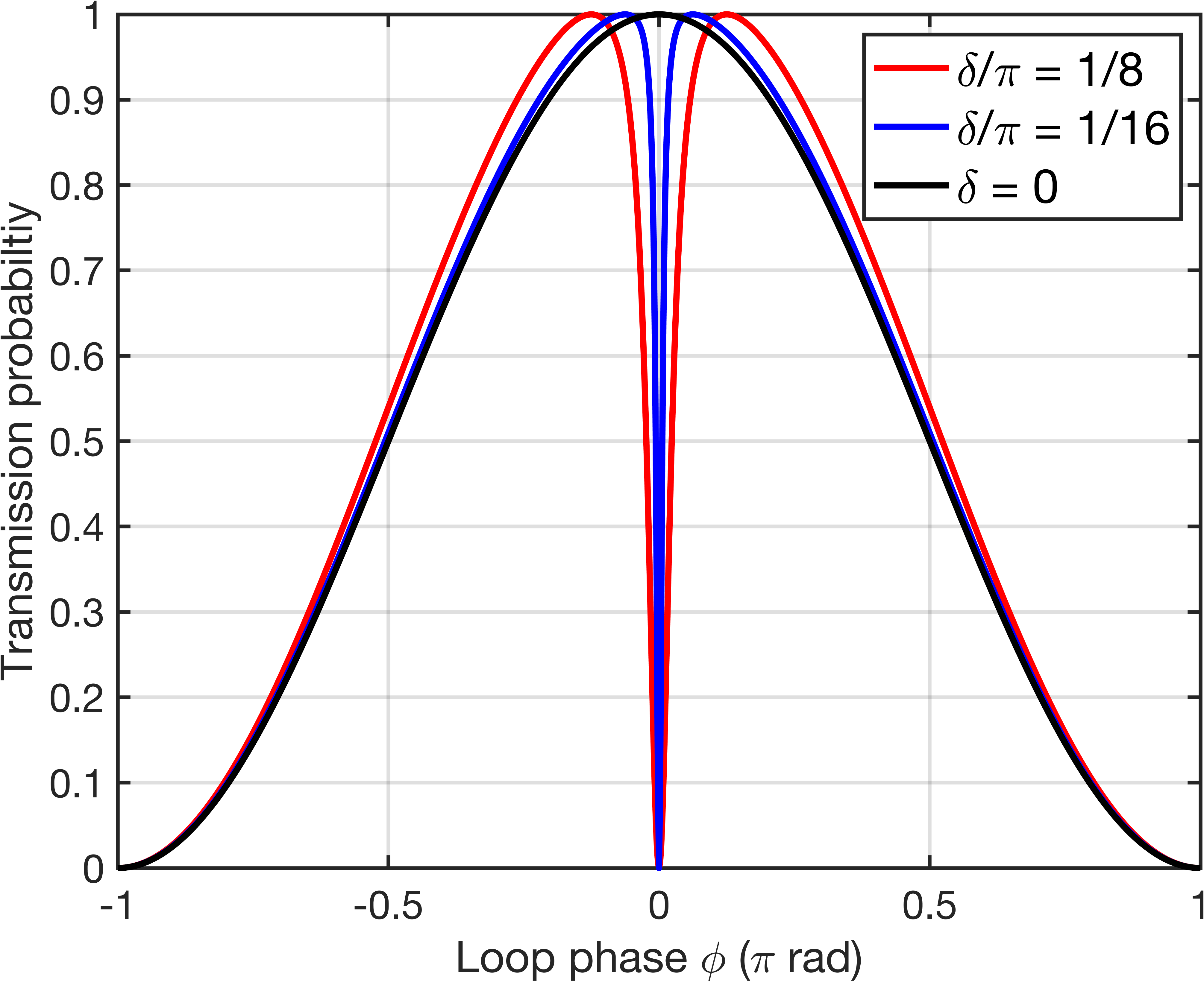}
    \caption{Transmission probability of the Grover-Sagnac interferometer vs. reciprocal loop phase $\phi$ for different values of the non-reciprocal phase $\delta$.}
    \label{fig:idealprobs}
\end{figure}

When $\delta$ is brought away from zero, the loop mode common-path symmetry is broken, so the subsequent overlap at the Grover coin admits back-reflections into the loop. In particular, the loop-mode portion of the state evolves like so: $(a_{\text{cw}}^\dagger + a_{\text{ccw}}^\dagger)\rightarrow e^{i\phi} (e^{i\delta}a_{\text{cw}}^\dagger + e^{-i\delta} a_{\text{ccw}}^\dagger)$. When these excitations are reintroduced to the Grover coin after traversing the loop (which swaps their incident ports), they become $e^{i\phi} (e^{i\delta} (a_1^\dagger + a_2^\dagger + a_{\text{cw}}^\dagger - a_{\text{ccw}}^\dagger) + e^{-i\delta} (a_1^\dagger + a_2^\dagger - a_{\text{cw}}^\dagger + a_{\text{ccw}}^\dagger) )/2 = e^{i\phi} (\cos\delta (a_1^\dagger + a_2^\dagger) + i\sin\delta (a_{\text{cw}}^\dagger - a_{\text{ccw}}^\dagger) ) $, and this excitation $(a_{\text{cw}}^\dagger - a_{\text{ccw}}^\dagger)$ is a cavity supermode, so it is easy to find its steady-state evolution. After a round-trip, its recursion is
\begin{subequations}
\begin{align}\label{eq:round-trip1}
(a_{\text{cw}}^\dagger - a_{\text{ccw}}^\dagger) &\rightarrow e^{i\phi}(e^{i\delta}a_{\text{cw}}^\dagger - e^{-i\delta}a_{\text{ccw}}^\dagger) \\ &\rightarrow e^{i\phi}( (e^{i\delta}(a_1^\dagger + a_2^\dagger + a_{\text{cw}}^\dagger - a_{\text{ccw}}^\dagger) \\ \notag &- (e^{-i\delta}(a_1^\dagger + a_2^\dagger - a_{\text{cw}}^\dagger + a_{\text{ccw}}^\dagger))/2 \\ &= e^{i\phi}(i\sin\delta (a_1^\dagger + a_2^\dagger) + \cos\delta (a_{\text{cw}}^\dagger - a_{\text{ccw}}^\dagger) ).\label{eq:round-trip2}
\end{align}
\end{subequations}
In the steady state, by definition, additional round-trips make no difference to the scattering state. The steady state can therefore be found by enforcing this condition, setting Eq. (\ref{eq:round-trip1}) equal to Eq. (\ref{eq:round-trip2}). The result is 
\begin{equation}
    (a_{\text{cw}}^\dagger - a_{\text{ccw}}^\dagger)_\text{ss} = \bigg (\frac{e^{i\phi} i\sin\delta}{1-e^{i\phi}\cos\delta} \bigg )(a_1^\dagger + a_2^\dagger).
\end{equation}
Combining this with the transient gives the final output state. Towards that end, we define 
\begin{equation}\label{eq:gamma}
    e^{i\gamma(\phi, \delta)} = e^{i\phi} \bigg ( \frac{\cos\delta - e^{i\phi}}{1 - e^{i\phi} \cos\delta}\bigg ), 
\end{equation}
so that the output can be written 
\begin{equation}
    |\psi_f^{(1)}\rangle = e^{i\gamma/2}(i\sin(\gamma/2) a_1^\dagger + \cos(\gamma/2)a_2^\dagger)
\end{equation}
with the scattering matrix generalizing to 
\begin{equation}\label{eq:nonstationary}
U_{\text{GS}}(\phi, \delta) =U_{\text{GS}}(\gamma(\phi, \delta)) = e^{i\gamma/2}
\begin{pmatrix}
    i\sin(\gamma/2) & \cos(\gamma/2)\\
    \cos(\gamma/2) & i\sin(\gamma/2)
\end{pmatrix}.
\end{equation}
Thus, introducing a nonzero $\delta$ is equivalent to replacing the reciprocal loop phase $\phi$ in Eq. (\ref{eq:stationary}) with the generalized phase shift $\gamma(\phi, \delta)$ in Eq. (\ref{eq:nonstationary}). The phase function may be expressed as 
\begin{equation}\label{eq:gamma2}
    \gamma(\phi, \delta) = \phi - \arctan \bigg (\frac{\sin\phi\sin^2\delta}{2\cos\delta - \cos\phi (1 + \cos^2\delta)} \bigg ).
\end{equation}
When $\delta = 0$, this family of functions takes on the well-defined value of $\gamma = \phi$ for all $\phi \neq 0$. However a general expansion of $\gamma(\phi)$ about $\delta = 0$ does not exist. As $\delta$ approaches 0 (or any other integer multiple of $\pi$), the slope $d\gamma/d\phi$ at the point $\phi = 0$ tends to infinity. The behavior is shown in Fig. \ref{fig:gammas}, where $\gamma(\phi)$ is plotted for different values of $\delta$. For small $\delta$, $\gamma(\phi)$ exhibits a region of very increased sensitivity about $\phi = 0$. This is highly a desirable property for detecting small $\delta$, since smaller $\delta > 0$ will generate increasingly larger changes in $\gamma$ that ultimately become sharp dips or peaks in the output probabilities. The correspondence between the slope $d\gamma/d\phi$ at $\phi = 0$ and $\delta$ is shown directly in Fig. \ref{fig:gammas}; this correspondence is mapped to the sharpness of the dip in the probability, as shown in Fig. \ref{fig:idealprobs}. For $\delta = 0$, the loop is single-pass and there is no dip, but for $\delta > 0$, the loop is multi-pass, causing a dip of full-width $2\delta = \Delta$ to appear in the sensitive $\phi$ region; the transmission peaks occur at $\phi = \pm \delta$, which can be seen by setting $\phi = \pm\delta$ in Eq. (\ref{eq:gamma}). This evaluates to $\gamma = 0$, so that the transmission probability is one at these points.

As shown in Fig. \ref{fig:gammas}, the nonlinear relationship between $\gamma$ and $\phi$ can be approximated by a linear one in the region near the inflection point $\phi = 0$. The slope of this linear region, $d\gamma/d\phi|_{\phi = 0}$ defines a scale factor for a linear increase in sensitivity. The phase response of the system will be more sensitive to changes in $\phi$ by this scale factor. This includes unwanted changes due to fluctuations, so the signal-to-noise ratio of the coherent input will not be affected. 

Away from $\phi = 0$ this linear approximation breaks down and the scale factor description no longer holds. The size of the region where this description is valid is the dynamic range. This may be rigorously defined by a value $\phi$ for which the description is valid to some small fixed error. As the value of $\delta$ moves away from $\pi/2$ towards $0$ or $\pi$, the slope scale factor continuously increases while the dynamic range and bandwidth continuously decrease, which is a standard feature of linear resonant systems. Anticipated numerical values of the slope and dynamic range can be readily extracted from Eq. (\ref{eq:gamma2}).

Insofar as the reflectance and/or transmittance peak extrema can be resolved in $\phi$, this device provides a direct readout for $\delta$ that does not require holding at a specific, precisely known bias point for $\phi$. More fundamentally it does not require global knowledge of $\phi$; unknown translations $\phi$ will not affect the width of the dips or peaks in the probability curves.

It is also interesting to observe that the symmetry-breaking caused by loop phase $\delta$ taking a nonzero value is associated with a topological change in the phase parameter space; the winding number of the generalized phase $\gamma(\phi, \delta)$ is one for $\delta = 0$ (dashed black curve in Fig. \ref{fig:gammas} but is two for all other fixed values of $\delta$. In accordance with this, we see the $2\pi$ periodicity of the probability curve for $\delta = 0$ (solid black curve in Fig. \ref{fig:idealprobs} jump into a $\pi$ periodicity in all other curves when $\delta$ is made nonzero.

In any resonant interferometer, the source coherence length must be larger than many round-trip lengths so that the paths corresponding to different round trips remain indistinguishable. The amplitudes of the indistinguishable paths are then coherently summed at the output. A monochromatic source assumption guarantees this condition is met for any loop size. Practical quasi-monochromatic sources always have a finite linewidth but even with inexpensive commercially available lasers, the linewidth can be so narrow that the coherence length exceeds several kilometers, allowing large loop sizes to be used. 

An estimate for the output state generated by a source with finite coherence length $\ell_c$ is readily found by computing the device output after a finite number round-trips have occurred \cite{PhysRevA.107.052615}, and summing only through the round-trip number where the total path length traversed is equal to the coherence length of the source. The error in this estimate is then approximately given by the magnitude of the next term in the series relative to the partial sum. This error is also decreased when losses are included in the loop, as the accumulated loss will rapidly drive higher-order terms in the series to zero. 

Choosing $\ell_c$ to be an order of magnitude larger than the round-trip distance $\ell$ is generally sufficient in practice render the effects of finite coherence length negligible. A more complete description of partial coherence, which would require a spectral analysis of the system, is beyond the scope of the current work and will be considered at a later time.

\section{Effect of loss and gain in the Sagnac loop}

The analyses above assume all underlying scatterers are unitary, including the loop phase shift, which can be represented by the following scattering matrix acting on the loop-mode basis $\{a_{\text{cw}}^\dagger, a_{\text{ccw}}^\dagger\}$:
\begin{equation}
\Phi(\phi, \delta) = e^{i\phi}
\begin{pmatrix}
    e^{i\delta} & 0\\
    0 & e^{-i\delta}
\end{pmatrix}.
\end{equation}

The global unitarity assumption ensures the output probabilities sum to one, but this will not occur in practice due to losses from sources such as scattering, absorption, a diverging beam, etc. Here we consider the effects of losses in each system to understand how the determination of $\delta$ is affected. 

Losses in the loop of either variant of the Sagnac interferometer can be taken into account using complex-valued phase parameters $u$ and $w$ in place of $\phi$ and $\delta$. We define $u = i\phi - \alpha$ and $w = i\delta - \beta$. $\alpha > 0$ represents reciprocal round-trip losses while $\alpha < 0$ indicates reciprocal gain. $\beta$ represents the presence of non-reciprocal gain or loss in the loop cavity. Thus, the amplitudes of the loop modes are scaled in magnitude by the factors $e^{\alpha + \beta}$ and $e^{\alpha - \beta}$ each round-trip. 

\subsection{Traditional Sagnac interferometer}

For losses which are purely reciprocal ($\beta = 0$), the traditional Sagnac interferometer is known to be robust. This follows from the output probabilities being insensitive to $\phi$. When $\phi$ is replaced with $\phi + i\alpha$ in Eq. (\ref{eq:sagnacS}), there is a degradation in total probability by a factor $e^{-2\alpha}$, but otherwise, there is no change in visibility or shape of the output probabilities. Such a decrease in signal will affect the readout of $\delta$, as unmeasured losses can be misinterpreted as a weaker $\delta$ than which is truly present. 

Finding the device output with the complex-valued quantities $u$ and $w$ generally leads to expressions with hyperbolic trigonometric functions with complex-valued arguments in place of the traditional ones. Accordingly, the scattering matrix in Eq. (\ref{eq:sagnacS}) generalizes to
\begin{equation}\label{eq:sagnacSgen}
    U_S(u, w) = e^u
    \begin{pmatrix}
        i\cosh w & \sinh w\\
            -\sinh w & i\cosh w
    \end{pmatrix}, 
\end{equation}
so that the transmission probability becomes 
\begin{equation}
    T(\delta) = |e^u \sinh w|^2 = e^{-2\alpha}(\sinh^2\beta \cos^2\delta + \cosh^2 \beta \sin^2 \delta).
\end{equation}
This remains sinusoidal in $\delta$ with reduced visibility whenever $\beta \neq 0$. The extreme values are $T(0) = e^{-2\alpha} \sinh^2 \beta$ and $T(\pi/2) = e^{-2\alpha} \cosh^2 \beta$, so the visibility $\mathcal{V}$ becomes
\begin{equation}
\mathcal{V} = \frac{\cosh^2 \beta - \sinh^2\beta}{\cosh^2 \beta + \sinh^2\beta} = \text{sech} (2\beta).
\end{equation}

\subsection{Grover-Sagnac interferometer}
With the possibility of gain and/or loss present, the function $e^{i\phi}$ will no longer be directly mapped to another parametrization of the unit circle $e^{i\gamma(\phi, \delta)}$. Instead, the generalization to Eq. (\ref{eq:gamma}) is the function $\Gamma: \mathbb{C}\times \mathbb{C} \rightarrow \mathbb{C}$, defined by
\begin{equation}\label{eq:Gamma}
    \Gamma(u, w) = e^u \frac{\cosh w - e^u}{1 - e^u \cosh w}.
\end{equation}
The output scattering amplitudes correspondingly become $r = (\Gamma - 1)/2$ and $t = r + 1$, so that the transmission probability is $T = |t|^2 = (|\Gamma|^2 + 2\text{Re}(\Gamma) + 1)/4$. 
\begin{figure}[ht]
    \centering
    \includegraphics[width=.5\linewidth]{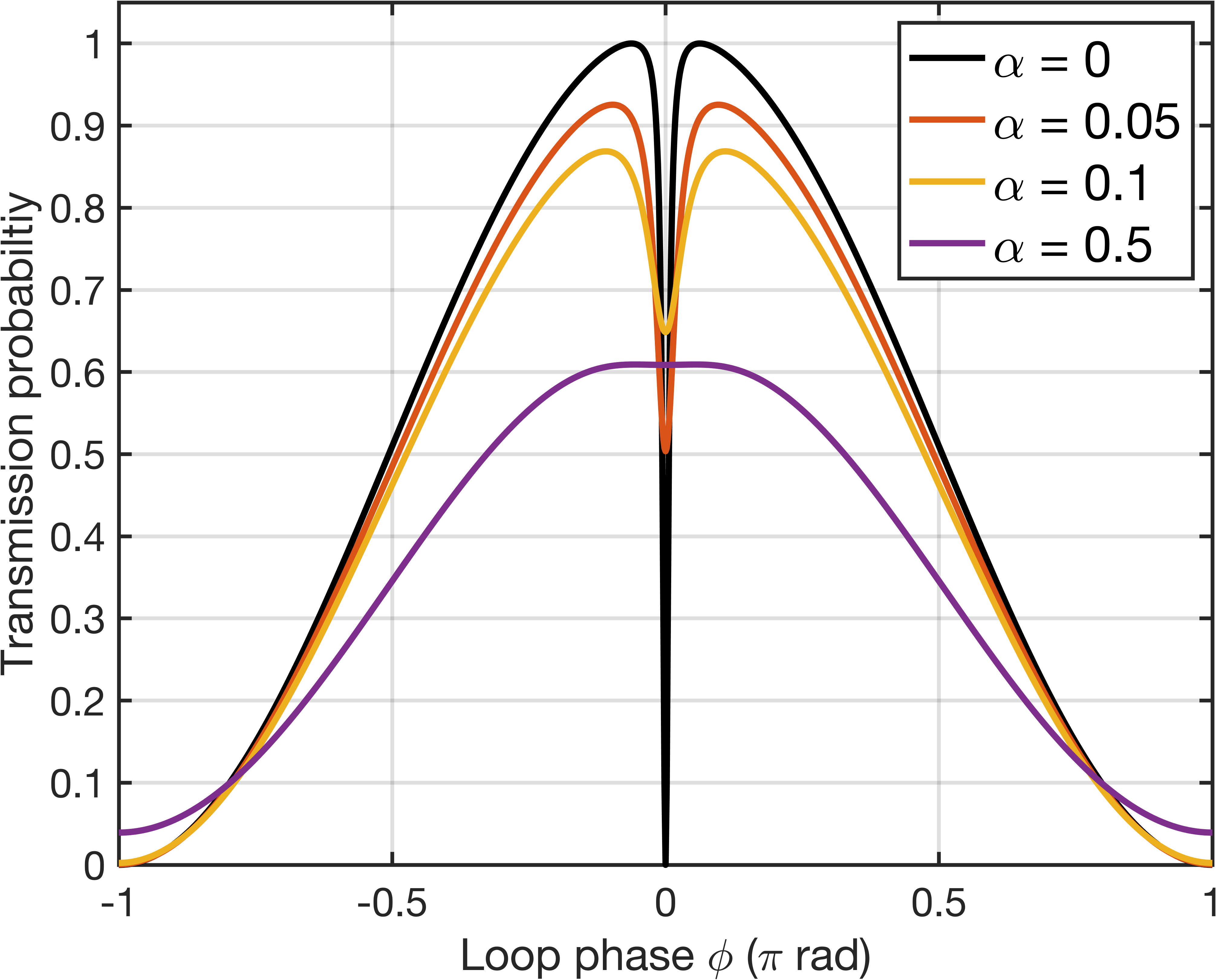}
    \caption{Transmission probability of the Grover-Sagnac interferometer vs. reciprocal loop phase $\phi$ for different values of the round-trip loss parameter $\alpha$. The other parameters are fixed with values $\delta = \pi/16$ and $\beta = 0$.}
    \label{fig:glprobs}
\end{figure}
\begin{figure}[ht]
    \centering
    \includegraphics[width=.5\linewidth]{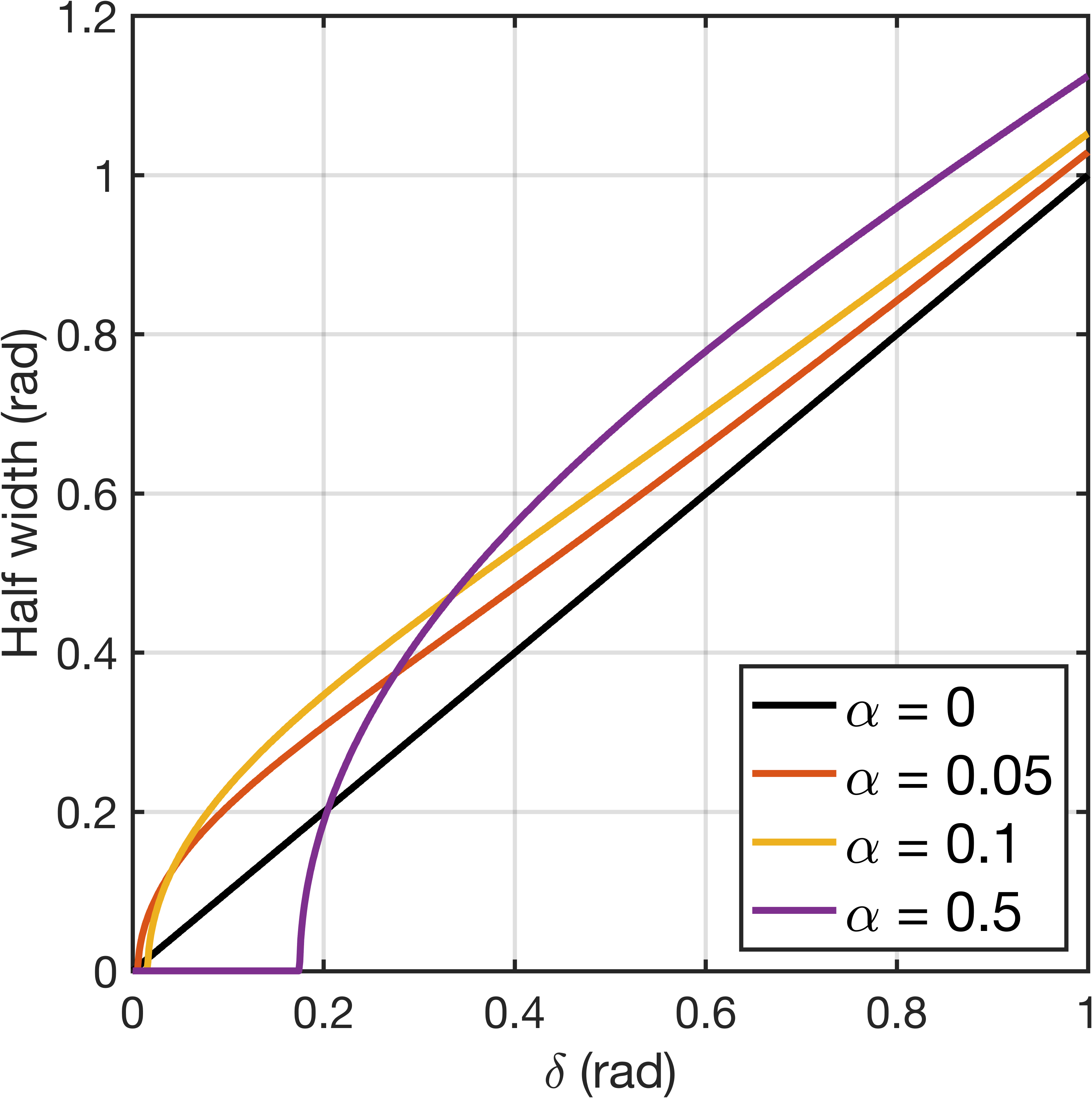}
    \caption{Half-width of transmission dip $\arg_{\phi > 0} \max T(\phi)$ for varying values of $\alpha$ with $\beta = 0$.}
    \label{fig:widths}
\end{figure}
Losses dampen the resonance shown in Fig. \ref{fig:idealprobs}, leading to a reduction in both visibility and width. These changes become more drastic when closer to the resonant point $(\phi, \delta) = (0, 0)$. This is illustrated in Fig. \ref{fig:glprobs}, which considers the effects of reciprocal loop loss alone: $T$ vs. $\phi$ is plotted several values of $\alpha$ with fixed $\delta = \pi/16$, $\beta = 0$. As the loss parameter $\alpha$ increases, the dip widens and eventually vanishes, limiting the minimum value of $\delta$ that can be extracted from width information alone. The half-width can be numerically obtained from Eq. (\ref{eq:Gamma}) and is shown for the same values of $\alpha$ and $\beta = 0$ in Fig. \ref{fig:widths}. For the unitary case, $\alpha = 0$, the splitting in dip location is linear in $\delta$. However, in the non-unitary cases when $\alpha$ is nonzero, below a certain threshold there is no dip. Nearby but greater than this threshold value of $\delta$, we see the splitting is comparatively sharper and nonlinear. Indeed the counter-propagating resonant loop modes are coupled by the Grover coin so they may form a second order exceptional point \cite{Wiersig:20} in a common-path configuration. This suggests that loss can offer a more sensitive, albeit more complicated readout for \textit{changes} in $\delta$ that is not present in a traditional Sagnac interferometer. These non-Hermitian dynamics will be further explored in a later work.

\begin{figure}[H]
    \centering
    \includegraphics[width=.5\linewidth]{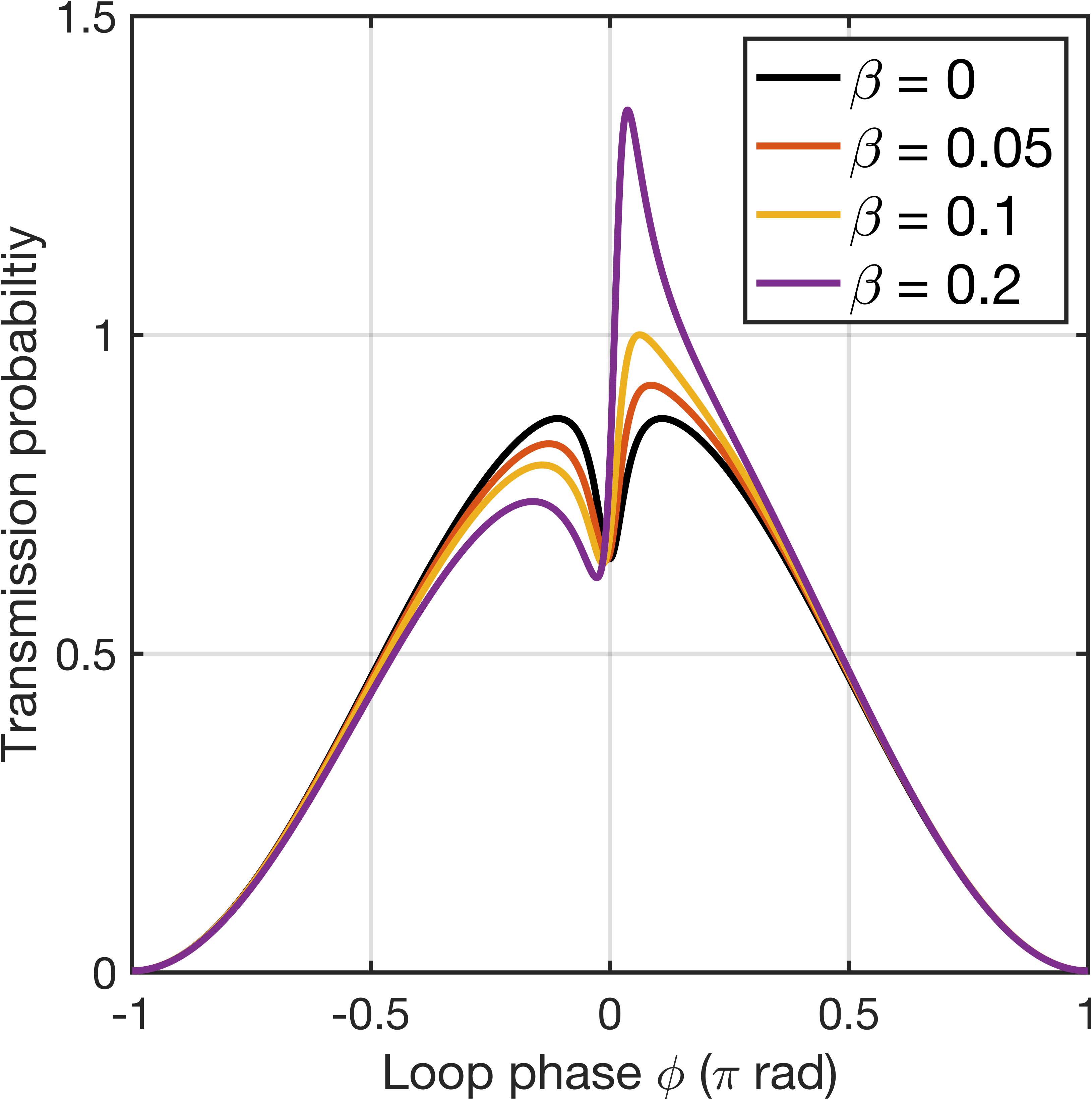}
    \caption{Transmission probability of the Grover-Sagnac interferometer for fixed reciprocal loss $\alpha = 0.1$ and $\delta = \pi/16$ for different values of the non-reciprocal loss/gain parameter $\beta$. Probabilities greater than one are a result of optical gain in the system.} 
    \label{fig:glprobs2}
\end{figure}

For $\beta$ nonzero, the resonance becomes skewed, as shown in Fig. \ref{fig:glprobs2}. While this complicates the task of inferring $\delta$, it's clear that information about $\beta$ lies in the transmission probability curve, for instance, in the difference in the heights of the maxima which form the dip. This opens up a potential to extract $\alpha, \beta$ and $\delta$ simultaneously by fitting the measured transmission to all parameters.
\section{Metrological applications}
In the above sections, the Grover-Sagnac interferometer's behavior was analyzed assuming a generic non-reciprocal phase $\Delta = 2\delta$ is applied in the loop. Here we consider some different uses the Grover-Sagnac interferometer might find with specific underlying $\Delta$. 

\begin{figure}[ht]
    \centering
    \includegraphics[width=\linewidth]{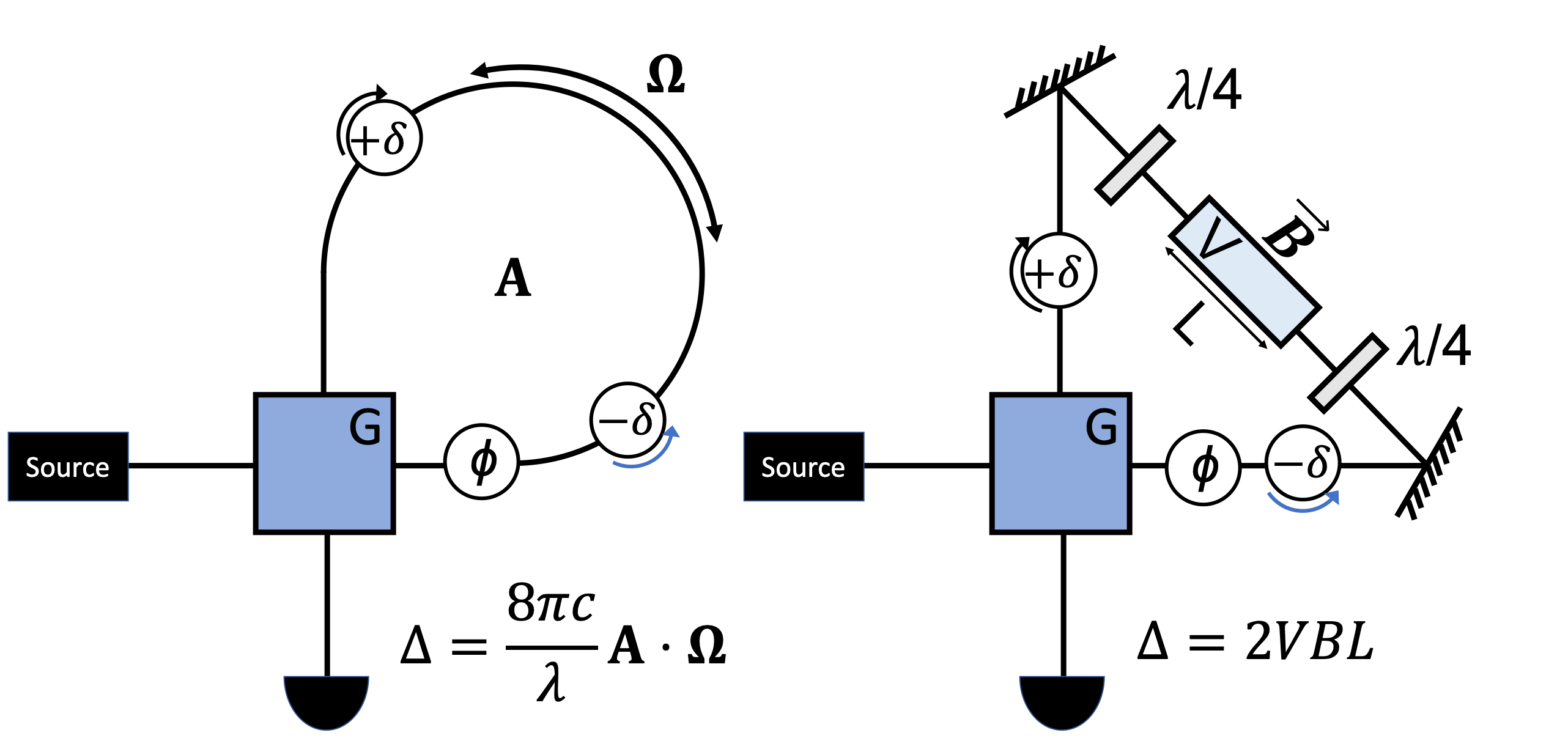}
    \caption{Schematic depiction of Grover-Sagnac interferometer in rotation (left) and magnetic field (right) sensing configurations.}
    \label{fig:schematics}
\end{figure}

The most obvious use is to measure rotations in conjunction with Eq. (\ref{eq:sagnac-phase}). An all-fiber Grover coin or one coupled to a large fiber loop would combine the traditional method of obtaining a large $\delta$ with the resonantly enhanced readout of $\delta$ to obtain a very sensitive apparatus for rotation measurements. A schematic of the device is shown in Fig. \ref{fig:schematics} (left). Care must be taken to balance round-trip losses, which scale with loop length, and the loop area. With adequately characterized loss, the system could be operated in the non-Hermitian regime to obtain enhanced sensitivity to changes in $\delta$ for sensing angular acceleration with increased resolution. 

Another envisioned application of the Grover-Sagnac interferometer is the sensing of weak magnetic fields. Both a conventional Sagnac and a Grover-Sagnac interferometer can be made to sense magnetic fields by inserting two quarter waveplates oriented at 45 degrees with respect to the horizontal plane into the Sagnac loop to convert between linear and circular polarization. A magneto-optic crystal, such as terbium gallium garnet (TGG), is then placed between the waveplates. An external magnetic field oriented along the optical axis of the crystal will induce a non-reciprocal phase shift equal in magnitude but opposite in sign for the two counter-propagating, circularly polarized beams. This phase shift is proportional to the strength of the magnetic field, the Verdet constant of the material (-143 rad/T/m for TGG for $\lambda =  633$ nm \cite{Barnes:92}), and the length of the crystal. A schematic for the proposed device is shown in Fig. \ref{fig:schematics} (right).

From the Verdet constant $V$ and length of the crystal $L$, one can obtain the magnetic field strength from the relation \cite{carothers.acs,hulme}
\begin{equation}
    B =\frac{\Delta}{2VL} = \frac{\delta}{VL}.
\end{equation}
The factor of two arises from the fact that the Verdet constant is specified in degrees of Faraday rotation of a linearly polarized beam propagating through a magneto-optic crystal under an applied magnetic field. The Faraday rotation angle is equal to \textit{half} of the phase delay between the two superimposed, orthogonal, circularly polarized beams.

Readout of the externally induced $\Delta = 2\delta$ can be conducted in a number of ways. One could bias at a particular value of $\phi$ and monitor the transmitted power for a direct readout of $\delta$. Alternatively, one may sweep the loop phase $\phi$ near zero to obtain an interferogram of the resonant transmission dip, and from this, extract a value of $\delta$. The second approach is unique to the Grover-Sagnac interferometer, since the conventional Sagnac performance is insensitive to changes in reciprocal phase. It allows in principle the detection of arbitrarily small magnetic fields but is limited practically by the Verdet constant and length of the magneto-optic crystal, the aggregate losses in the loop, and the granularity of phase scanning resolution. Other readout techniques might be imagined, such as a system which images a fringe pattern. Instead of a standard shift in fringe position, changing $\delta$ in a Grover-Sagnac system would be expected to generate fine breaks inside the main fringes, corresponding to the dips or peaks in transmission or reflection. The ratio of the break width to that of the main fringe would reveal $\Delta$.

\section{Conclusions}
Despite originating from the same scattering graph topology, the Grover-Sagnac interferometer behaves very differently than its conventional predecessor. This is due to the interplay of higher-dimensional modal interference, especially that which results from the directionally-unbiased coin coherently coupling into back reflections. The simultaneous, symmetric excitation of both loop modes creates a common-path coupled cavity system leading to the resonant response in $\delta$. The presence of a coupling between the counter-propagating modes of the common-path ring cavity is a direct consequence of using a directionally unbiased central scatterer.

The generalization of the internal phase $\phi$ to $\gamma(\phi)$ that occurs when $\delta$ is made nonzero occurs in other resonant optical systems \cite{PhysRevA.109.053508, GTI, 10.5555/553457} albeit typically with only reciprocal phases. The general phase amplification principle, which uses entirely linear-optical scatterers, can be useful in many contexts for converting weak input phase changes to effectively large ones.

The dip readout of the Grover-Sagnac system is somewhat antithetical to the traditional readout approach. In the conventional Sagnac systems, $\delta$ is registered by measuring small changes in power, so that the smallest $\delta$ measurable and thus the system resolution is ultimately limited by the granularity of the detector system with respect to changes in output optical energy. In the Grover-Sagnac, however, the limiting factor is the resolution of the spanned loop phase $\phi$, as this must be finer than the transmission dip in order to resolve it. Therefore this apparatus must be able to impart, rather than detect, increasingly fine changes in the inputted energy in order to increase resolution. This viewpoint of course assumes no loss, which will degrade the readout resolution in both systems.

Beyond the loss of optical signal, fabrication errors can additionally result in passive coins scattering with undesired power ratios. For coins such as the beam-splitter or Grover multiport, where the intended splitting ratios are equal, these deviations are another form of symmetry-breaking dual to the presence of a nonzero $\delta$. Indeed, fabrication errors will usually induce losses in an asymmetric fashion, so that losses and scattering imbalances appear jointly, making the general situation more complicated. A more detailed analysis of the non-Hermitian dynamics and symmetry-breaking of the coin action is forthcoming.

Some Sagnac interferometer variants have been proposed and demonstrated for applications centered around the reciprocal phase $\phi$. Because non-reciprocal portions of the loop phase are relatively less common in nature, systems involving $\phi$ that are controlled by this phase $\delta$ are generally more isolated from environmental fluctuations. For example, this is the underlying principle of the Sagnac loop mirror, and might be extended to select a more stable phase amplification constant $d\gamma/d\phi$ than in a fully reciprocal phase sensor such as a Grover-Michelson interferometer or ring-loaded Mach-Zehnder interferometer. The intrinsic weakness of many sources of non-reciprocal phases would allow the phase amplification constant to be controlled with increased precision. For example, a magneto-optic material in the loop may be controlled by an external magnetic field. This allows the Grover-Sagnac interferometer to double as a useful phase-amplified readout of a generic phase $\phi$ with tunable but stable phase gain, which may improve systems for refractive index determination and phase contrast imaging with tunable magnification.

Overall, the system is another instance illustrating that symmetric scatterers with increased dimensionality can improve an interferometric system. Other unbiased scattering coins may provide other useful benefits.

\section*{Funding.}
Air Force Office of Scientific Research MURI award number FA9550-22-1-0312.
\section*{Disclosures.}
None to report.
\section*{Data availability.}
All data generated for this article is available upon reasonable request to the authors.
\bibliography{refs}

\begin{thebibliography}{10}
\newcommand{\enquote}[1]{``#1''}

\bibitem{PhysRevD.38.2317}
B.~J. Meers, \enquote{Recycling in laser-interferometric gravitational-wave
  detectors,} {\protect\JournalTitle{Phys. Rev. D}} \textbf{38}, 2317--2326
  (1988).

\bibitem{Fritschel:92}
P.~Fritschel, D.~Shoemaker, and R.~Weiss, \enquote{Demonstration of light
  recycling in a michelson interferometer with fabry--perot cavities,}
  {\protect\JournalTitle{Appl. Opt.}} \textbf{31}, 1412--1418 (1992).

\bibitem{PhysRevLett.66.1391}
K.~A. Strain and B.~J. Meers, \enquote{Experimental demonstration of dual
  recycling for interferometric gravitational-wave detectors,}
  {\protect\JournalTitle{Phys. Rev. Lett.}} \textbf{66}, 1391--1394 (1991).

\bibitem{LIGO}
A.~Abramovici, W.~E. Althouse, R.~W.~P. Drever, Y.~Gürsel, S.~Kawamura, F.~J.
  Raab, D.~Shoemaker, L.~Sievers, R.~E. Spero, K.~S. Thorne, R.~E. Vogt,
  R.~Weiss, S.~E. Whitcomb, and M.~E. Zucker, \enquote{Ligo: The laser
  interferometer gravitational-wave observatory,}
  {\protect\JournalTitle{Science}} \textbf{256}, 325--333 (1992).

\bibitem{Gray:98}
M.~B. Gray, A.~J. Stevenson, H.-A. Bachor, and D.~E. McClelland,
  \enquote{Broadband and tuned signal recycling with a simple michelson
  interferometer,} {\protect\JournalTitle{Appl. Opt.}} \textbf{37}, 5886--5893
  (1998).

\bibitem{PhysRevA.107.052615}
C.~R. Schwarze, D.~S. Simon, and A.~V. Sergienko, \enquote{Enhanced-sensitivity
  interferometry with phase-sensitive unbiased multiports,}
  {\protect\JournalTitle{Phys. Rev. A}} \textbf{107}, 052615 (2023).

\bibitem{Weihs:96}
G.~Weihs, M.~Reck, H.~Weinfurter, and A.~Zeilinger, \enquote{All-fiber
  three-path mach--zehnder interferometer,} {\protect\JournalTitle{Opt. Lett.}}
  \textbf{21}, 302--304 (1996).

\bibitem{PhysRevA.106.033706}
D.~S. Simon, C.~R. Schwarze, and A.~V. Sergienko, \enquote{Interferometry and
  higher-dimensional phase measurements using directionally unbiased linear
  optics,} {\protect\JournalTitle{Phys. Rev. A}} \textbf{106}, 033706 (2022).

\bibitem{PhysRevA.110.023527}
C.~R. Schwarze, A.~D. Manni, D.~S. Simon, and A.~V. Sergienko,
  \enquote{Single-photon description of the lossless optical y coupler,}
  {\protect\JournalTitle{Phys. Rev. A}} \textbf{110}, 023527 (2024).

\bibitem{Clements:16}
W.~R. Clements, P.~C. Humphreys, B.~J. Metcalf, W.~S. Kolthammer, and I.~A.
  Walmsley, \enquote{Optimal design for universal multiport interferometers,}
  {\protect\JournalTitle{Optica}} \textbf{3}, 1460--1465 (2016).

\bibitem{Schwarze:24}
C.~R. Schwarze, D.~S. Simon, A.~D. Manni, A.~Ndao, and A.~V. Sergienko,
  \enquote{Experimental demonstration of a grover-michelson interferometer,}
  {\protect\JournalTitle{Opt. Express}} \textbf{32}, 34116--34127 (2024).

\bibitem{Shaddock:98}
D.~A. Shaddock, M.~B. Gray, and D.~E. McClelland, \enquote{Experimental
  demonstration of resonant sideband extraction in a sagnac interferometer,}
  {\protect\JournalTitle{Appl. Opt.}} \textbf{37}, 7995--8001 (1998).

\bibitem{Liu:23}
S.~Liu, S.~Zhao, Q.~Liu, S.~Jiang, W.~Jin, and Z.~He,
  \enquote{Ultrahigh-resolution and ultra-simple fiber-optic sensor with
  resonant sagnac interferometer,} {\protect\JournalTitle{Opt. Lett.}}
  \textbf{48}, 3543--3546 (2023).

\bibitem{RevModPhys.39.475}
E.~J. Post, \enquote{Sagnac effect,} {\protect\JournalTitle{Rev. Mod. Phys.}}
  \textbf{39}, 475--493 (1967).

\bibitem{MICHELSON1925}
A.~A. Michelson and H.~G. Gale, \enquote{The effect of the earth's rotation on
  the velocity of light,} {\protect\JournalTitle{Nature}} \textbf{115},
  566--566 (1925).

\bibitem{Usman24_2}
A.~Usman, A.~Bhatranand, Y.~Jiraraksopakun, K.~S. Muhammad, and P.~Buranasiri,
  \enquote{Nio thickness measurement using a rectangular-type sagnac
  interferometer as the material transport layer in a perovskite solar cell,}
  {\protect\JournalTitle{Appl. Opt.}} \textbf{63}, 2868--2875 (2024).

\bibitem{Usman_2022}
A.~Usman, Y.~Jiraraksopakun, R.~Kaewon, C.~Pawong, and A.~Bhatranand,
  \enquote{The comparison of multi-stepping algorithms for real-time thickness
  measurement of transparent thin films using polarization settings,}
  {\protect\JournalTitle{Laser Physics}} \textbf{32}, 125401 (2022).

\bibitem{10597582}
R.~Shi, H.~Chen, X.~Fan, C.~Liu, H.~Li, Z.~Gao, S.~Zhang, M.~Gu, L.~Li,
  Y.~Zheng, and S.~Li, \enquote{Highly sensitive strain and temperature sensors
  based on vernier effect in cascaded mach–zehnder interferometer and sagnac
  interferometer,} {\protect\JournalTitle{IEEE Transactions on Instrumentation
  and Measurement}} \textbf{73}, 1--10 (2024).

\bibitem{Usman24_1}
A.~Usman, A.~Bhatranand, Y.~Jiraraksopakun, K.~S. Muhammad, and P.~Buranasiri,
  \enquote{Phase-shifting determination and pattern recognition using a
  modified sagnac interferometer with multiple reflections,}
  {\protect\JournalTitle{Appl. Opt.}} \textbf{63}, 1135--1143 (2024).

\bibitem{10.1063/5.0123236}
H.~Arianfard, S.~Juodkazis, D.~J. Moss, and J.~Wu, \enquote{{Sagnac
  interference in integrated photonics},} {\protect\JournalTitle{Applied
  Physics Reviews}} \textbf{10}, 011309 (2023).

\bibitem{Wiersig:20}
J.~Wiersig, \enquote{Review of exceptional point-based sensors,}
  {\protect\JournalTitle{Photon. Res.}} \textbf{8}, 1457--1467 (2020).

\bibitem{Barnes:92}
N.~P. Barnes and L.~B. Petway, \enquote{Variation of the verdet constant with
  temperature of terbium gallium garnet,} {\protect\JournalTitle{J. Opt. Soc.
  Am. B}} \textbf{9}, 1912--1915 (1992).

\bibitem{carothers.acs}
K.~J. Carothers, R.~A. Norwood, and J.~Pyun, \enquote{High verdet constant
  materials for magneto-optical faraday rotation: A review,}
  {\protect\JournalTitle{Chemistry of Materials}} \textbf{34}, 2531--2544
  (2022).

\bibitem{hulme}
H.~R. Hulme and R.~H. Fowler, \enquote{The faraday effect in ferromagnetics,}
  {\protect\JournalTitle{Proceedings of the Royal Society of London. Series A,
  Containing Papers of a Mathematical and Physical Character}} \textbf{135},
  237--257 (1932).

\bibitem{PhysRevA.109.053508}
C.~R. Schwarze, D.~S. Simon, A.~Ndao, and A.~V. Sergienko, \enquote{Tunable
  linear-optical phase amplification,} {\protect\JournalTitle{Phys. Rev. A}}
  \textbf{109}, 053508 (2024).

\bibitem{GTI}
F.~Gires and P.~Tournois, \enquote{Interferometre utilisable pour la
  compression dimpulsions lumineuses modulees en frequence,}
  {\protect\JournalTitle{Comptes Rendus Hebdomadaires Des Seances De L Academie
  Des Sciences}} \textbf{258}, 6112 (1964).

\bibitem{10.5555/553457}
C.~K. Madsen and J.~Zhao, \emph{Optical Filter Design and Analysis: A Signal
  Processing Approach} (John Wiley \& Sons, Inc., USA, 1999), 1st ed.

\end{thebibliography}

\end{document}